\documentclass[12pt,a4paper]{article}

\usepackage{graphicx,graphics,color,epsfig}
\usepackage{latexsym}
\usepackage{amsmath,amsfonts,amssymb}
\usepackage{verbatim}

\begin{document}

\begin{center}
\Large {\bf Equations for fields with additional fundamental physical constants}
\vskip 8mm
\large
V. V. Khruschov
\vskip 5mm
\large
{{\it Center of Gravitation and Fundamental Metrology, VNIIMS, 
		Ozyornaya ul. 46, Moscow 119361, Russia }} \\
\end{center}

\vskip 6mm

\begin{abstract}
\noindent
The author investigates the general Lie algebra of operators of coordinates, momenta, and Lorentz group generators, which can be used in quantum gravity, theories with generalized uncertainty principle, double and triple relativity and theories with noncommutative space-time and momentum spaces. The structure constants of this algebra depend on the constants {\it c} and {\it h}, as well as additional constants with dimensions of action ({\it H}), length ({\it L}), and mass ({\it M}). In the limiting case of infinite {\it H}, {\it L}, and {\it M}, the algebra goes into that of operators of the canonical quantum theory in Minkowski space-time. Some representations of this algebra and equations for generalized fields depending on additional fundamental physical constants are given. 
\end{abstract}
\vspace*{6pt}

\bigskip

\vskip 8mm

\section{Introduction}
\label{Section_Introduction}
Spacetime which is the arena of operations both for the Standard Model of strong and electroweak interactions and the theory of General Relativity (GR) has different meanings 
in the frameworks of these theories. In GR space-time is a classical manifold whereas in the Standard Model  space-time coordinates have nontrivial commutation relations with momentum operators. It is expected that  space-time  will be the same in nature in a unified theory of quantum gravity and strong and electroweak interactions 
\cite{dopl,gara,amel}. At transition from a microscopical space-time to  a classical one 
a decoherence  occurrs through the presence of an environment and quantum or classical gravitation \cite{arza}. Moreover in this case the microscopic properties of space-time and momentum operators will became more complicate and their commutation relations among themselves will change (see, e.g. \cite{toll}). This leads to theories with a generalized uncertainty principle  \cite{magg,kemp,adle,pali}, double \cite{amel2,brun,magu} and triple special relativity  \cite{kowa}, which are admissible in the general case for description of  microscopic  curved and noncommutative space-time  \cite{melj}.

Much remains to be done in a generalization of the canonical quantum theory
in the flat Minkowski spacetime.
For example, usually it is considered that  a scale of appearance of quantum gravity effects  is given by the Plank mass. However the possibility exists of a considerable reduction of this scale, for instance, in a theory of strings\cite{arza} or in semiclassical models for replacement of singularities in GR solutions \cite{bron}. So it is instructive to consider models which allow for varied values of additional fundamental physical constants. 
In this paper we develop the generalized Snyder model  with  
additional varied constants with dimensions of action ({\it H}), length ({\it L}), and mass ({\it M}).
 
The first model of this type had its origin in  the work of Snyder \cite{Snyder}  and subsequently was elaborated in the Refs. \cite{gol,kady}  as a theory with a minimal length for a removal of ultraviolet divergences in the canonical quantum electrodynamics. Such divergences originate from a multiplication of quantities in close neighbouring spacetime points, therefore Heisenberg assumed that coordinates can be noncommutative \cite{jack}. It is important that  a minimal length was introduced  by Snyder in the Lorentz invariant manner.
However, in the Yang's paper \cite{Yang}  it was observed  that  the interchangeability among coordinates and momenta, proposed by Born \cite{Born1,Born2},
was lost in the Snyder's approach. Thus, it was necessary to replace the Poincar{\'e} group by the semi-direct product of the Lorentz group and the Heisenberg group and  to restore  the  interchangeability. Then two new constants {\it L} and {\it M} with dimensions of length and mass appeared in the theory \cite{Yang,lez}. More general model with three additional constants with dimensions of action ({\it H}), length ({\it L}), and mass ({\it M}) with varied values was introduced in papers \cite{lekh,khle}. Subsequently various models, depending on these constants,  were considered in Refs. \cite{toll,iz92,iz94,mend,ch1,ch2}.
 
In this paper the general Lie algebra with additional fundamental physical constants $L$, $M$, and $H$ is considered and some representations and equations in representation space of this algebra are presented.  In Sec. 2, we describe the $HLM$ algebra  which generalizes the semi-direct sum of the Lorentz algebra and the Heisenberg algebra. In Sec. 3  the equation  for a scalar $HLM$ generalized field is presented. 
 In Sec. 4, we outline the case of $LM$ generalized fields, which can be used for description of particles participated in  interactions conserving the $P$, $C$, and $T$ parities. In Sec. 5, we 
present the equations  for spinor $LM$ generalized fields with  conservation and  non-conservation of the $P$ parity. We  draw brief conclusions in Sec. 6.

\section{Lie algebras of generalized symmetries in quantum phase space}
\label{Section_anomal}

Below we consider  groups which generalize the semi-direct product of the Lorentz group and the Heisenberg one  in an eight-dimensional phase space. We  restrict our consideration to the determination of a group structure near the identity element of a group, so it is sufficient to investigate admissible generalized Lie algebras. 

The canonical commutation relations among the quantum theory observables in  Minkowski space-time   are well known \cite{Bo_Lo_To}:
\[
[x_{i},x_{j}]=[p_{i},p_{j}]=0,\quad \lbrack p_{i},x_{j}]=i\hbar
g_{ij}I,
\]
\[
[p_{i},I]=[x_{i},I]=[F_{ij},I]=0,
\]
\[
[F_{ij},F_{kl}]=i\hbar
(g_{jk}F_{il}-g_{ik}F_{jl}+g_{il}F_{jk}-g_{jl}F_{ik}),
\]
\begin{equation}
[F_{ij},p_{k}]=i\hbar (g_{jk}p_{i}-g_{ik}p_{j}),
\label{standart}
\end{equation}
\[
[F_{ij},x_{k}]=i\hbar (g_{jk}x_{i}-g_{ik}x_{j}),
\]
\noindent The relations (\ref{standart}) are written in the system of units, where 
the speed of light  $ c = 1$, $x_{i}$ are coordinate operators, $p_{i}$ are momentum operators and $I$ is identity operator, while $F_{ij}$ are generators of the Lorentz group; $i,j,k,l = 0,1,2,3$. 

We define spacetime points in the Minkowski space in the terms of Lie algebra representation theory  as eigenvalues of operators $x_{i}$, realized in a certain representation of the algebra (\ref{standart}). For example, fields in a $x-$representation are given as follows 
$\psi^S_{\alpha}(x)$, where $x=\{x_0,x_1,x_2,x_3\}$ are eigenvalues of coordinate operators $x_{i}$, $\alpha$ are discrete indexes for some finite-dimensional representation of the spin  operators $S_{ij}=F_{ij}-x_{i}p_{j}+p_{i}x_{j}$. 

One can consider possible generalizations of the Lie algebra (\ref{standart})
 under   fulfilment of the following conditions  \cite{lekh,khle}:
a) the generalized algebra should be a Lie algebra,
b) the dimensionality of the generalized algebra  and the physical dimensions of operators contained in should be the same as in the algebra  (\ref{standart}),
c) the generalized algebra should contain the Lorentz algebra as its subalgebra, and commutation relations of the Lorentz subalgebra with other generators should be the same as in the initial algebra. 

The procedure of generalization of the algebra  (\ref{standart}) described above may be named as a relativistic or a Lorentz-invariant deformation of the algebra (\ref{standart}) because the property of the Lorentz symmetry is conserved as the fundamental law of nature. However in some cases the canonical Poincar{\'e} invariance can be violated. 
The Lie algebra with the generators $F_{ij}$, $p_{i}$, $x_{i}$, and $I$, that is the maximal generalization of the algebra (\ref{standart}) under the conditions written above, has the following form \cite{lekh,khle}: 
\[
[F_{ij},F_{kl}]=if(g_{jk}F_{il}-g_{ik}F_{jl}+g_{il}F_{jk}-g_{jl}F_{ik}),
\]%
\[
[p_{i},x_{j}]=if(g_{ij}I+\frac{F_{ij}}{H}),
\]%
\[
[p_{i},p_{j}]=\frac{if}{L^{2}}F_{ij},
\]%
\[
[x_{i},x_{j}]=\frac{if}{M^{2}}F_{ij},
\]%
\begin{equation}
[p_{i},I]=if(\frac{x_{i}}{L^{2}}-\frac{p_{i}}{H}),
\label{finalg}
\end{equation}%
\[
[x_{i},I]=if(\frac{x_{i}}{H}-\frac{p_{i}}{M^{2}}),
\]%
\[
[F_{ij},p_{k}]=if(g_{jk}p_{i}-g_{ik}p_{j}),
\]%
\[
[F_{ij},x_{k}]=if(g_{jk}x_{i}-g_{ik}x_{j}),
\]%
\[
[F_{ij},I]=0.
\]

The commutation relations of the algebra  (\ref{finalg}) depend on four dimensional parameters:  $L$ with dimensions of length,  $M$ with dimensions of mass,  $H$ and $f$ with  dimensions of action ( $M$ and $L$ take real values as well as pure imaginary ones, $c=1$  in the system of units being used).
 
 In the limiting case, when  
 $M$, $L$, and $H$ become infinitely large, the commutation relations 
 (\ref{finalg}) turn into the commutation relations of the canonical algebra (\ref{standart}) providing $f=\hbar $. 
A more complicated case is also possible, when $f$ is some function of the 
parameters $L$, $M$, and $H$. This function $f(L,M,H)$ must tend to $\hbar$  for 
accordance with the conventional commutation relations at 
 $L\rightarrow \infty $, $M\rightarrow \infty $, and 
 $H\rightarrow \infty $.

Otherwise, when $f=\hbar $, but $M$, $L$, and $H$ 
have different limits, the commutation relations  transfer to the relations   considered earlier in the papers of Snyder   \cite{Snyder}, Yang \cite{Yang}, Golfand \cite{gol}, Kadyshevsky  \cite{kady}, and Leznov \cite{lez}, namely,
$H\rightarrow \infty $, $L\rightarrow \infty $ - the relativistic quantum theory with noncommutative coordinates;
$H\rightarrow \infty$, $M\rightarrow \infty$ - the relativistic quantum theory with noncommutative momenta;
$H\rightarrow \infty $ - the relativistic quantum theory with noncommutative coordinates and momenta.

The system of the commutation relations (\ref{finalg}) specifies some class of Lie algebras which consist of semisimple algebras as well as general-type algebras. After calculation of the Killing-Cartan form, the condition of semisimplicity for the algebras (\ref{finalg}) may be written as \cite{khle}
\begin{equation}
\frac{f^{2}(M^{2}L^{2}-H^{2})}{H^{2}M^{2}L^{2}}\neq 0.
\label{killing}
\end{equation}

\noindent To define all semisimple algebras of the type (\ref{finalg}),
it is convenient to perform the following transformation of generators 
 $p_{i}$, $x_{i}$, and $I$ \footnote{We keep the former notations for the operators  $p_i$, $x_i$, and  $I$, however now  the $p_i$, $x_i$, and  $I$ symbols mean the operators of the generalized momenta, coordinates and accordingly, the generalized 'identity' operator. }
\[
F_{i5}=Bx_{i}+Dp_{i},
\]
\begin{equation}
 F_{i6}=Ex_{i}+Gp_{i},\quad F_{56}=AI,
\label{lin_tr}
\end{equation}

\noindent then, under condition (\ref{killing}), one may obtain the  commutation relations for the algebras of pseudo-orthogonal groups 
$O(3,3)$, $O(2,4)$, and $O(1,5)$. 
These algebras correspond to specific values of the parameters 
$M^{2}$, $L^{2}$, and  $H^{2}$ \cite{khle}.

For 
\begin{equation}
 H^{2}=M^{2}L^{2}, \qquad M^{2}>0, \qquad L^{2}>0
\end{equation}   
\noindent the  $o(1,5)$ algebra degenerates into a semi-direct product of the 
 $o(1,4)$ algebra and the algebra of 5-dimensional translations, while for 
\begin{equation}
 H^{2}=M^{2}L^{2}, \qquad M^{2}<0, \qquad L^{2}<0
\end{equation}   
\noindent the  $o(3,3)$ algebra degenerates into a semi-direct product of the   $o(2,3)$ algebra and the algebra of 5-dimensional translations.
The special importance have the following values of parameters:  $L^2=0$, or $M^2=0$, or  
$H^2=L^2M^2$, because passing through these values mutates a type of an  algebra
under consideration.

\section{Some representations of $HLM$ algebras. 
Equation for a generalized scalar field}
\label{Section_OscillationModel}

Irreducible representations of the algebras (\ref{finalg}) are specified with  eigenvalues of Casimir operators. For the considered real simple algebras  the Casimir operators have the known forms in terms of the generators $F_{ij}$,  
$i,j=0,1,...,5 $ of the pseudoorthogonal groups in six-dimensional spaces
\cite{Jeloben,xelgas}: 
\[
C_1=\epsilon_{ijklmn}F^{ij}F^{kl}F^{mn}, \quad C_2=F_{ij}F^{ij},
\]
\begin{equation}
C_3=(\epsilon_{ijklmn}F^{kl}F^{mn})^2
\label{cas}
\end{equation}

\noindent However the Casimir operators should be expressed through the operators   $p_i$, $x_i$,
$F_{ij}$, $i,j=0,...,3$, and $I$ to be used in physical applications. 

Note that the general algebra (\ref{finalg}) with two constants with the dimension of action, 
namely $f$ and  $H$, is non invariant with  respect to  $T$ and 
$C$ transformations \cite{lekh,khle}. In the framework of the orthodox theory the Planck constant is the single odd constant with respect to  $T$  transformation (the speed of light enters in the commutation relations in the second power). So the transformation to conjugate or transposed operators leads to the recovery of the $T$ invariance, however it is impossible in the theory with two constants  $f$ and  $H$. Along the same lines one may obtain $C$ non-invariance of the system (\ref{finalg}), since the quantities with dimension of mass change their signs after replacement of particles with antiparticles. 
 The $CT$ transformation does not change  $f$ and $H$, so the system (\ref{finalg}) is invariant under $CT$ and $P$ transformations.

At $M\to\infty$, $L\to\infty$ the commutativity of coordinates among themselves, as well as momenta, restore.  However, the commutation relations among   $I$, $p_i$, $x_j$, and $x_j$, $p_i$ keep non-trivial. In this case it is not difficult to find a six dimensional representation of the algebra  (\ref{finalg}) with the help of the simplest matrices $e^i_j$, which contain unity at the intersection of the row  $i$ and the column $j$. Accomplish these ends it is needed to form the matrices 
$M^i_j=$$-e^i_j+e^j_i$ for $i,j=1,...,4$, $i<j$, and also for $i=0$, $j=5$, as well as the matrices 
 $N^i_j=$$e^i_j+e^j_i$ for $i=0$, $j=1,...,4$, and also for  
$j=5$, $i=1,...,4$. Then linear combinations with real coefficients of the matrices  $M^i_j$ and $N^i_j$ realize the real six dimensional representation of the initial algebra \cite{iz94}.

Let us give the infinite dimensional representation of the generators $F_{ij}$,
 $p_i$, $x_i$, and $I$ with the help of differential operators in the
 $\xi-$representation at $M\to\infty$, $L\to\infty$ \cite{iz94} (in the formulae below $a$ is a free 
parameter). This representation is the most similar to a representation of the canonical quantum theory and 
is dependent only on the $H$ constant.
\[
p_i=i\hbar\frac{\partial}{\partial\xi^i}, \qquad
I=a-i\hbar\frac{\xi^m}{H}\frac{\partial}{\partial\xi^m},
\]
\begin{equation}
F_{ij}=i\hbar(\xi_i\frac{\partial}{\partial\xi^j}-
\xi_j\frac{\partial}{\partial\xi^i}), 
\label{ksi}
\end{equation}
\[
x_i=a\xi_i-i\hbar(\frac{\xi_i\xi^m}{H}\frac{\partial}{\partial\xi^m}-
\frac{\xi^2}{2H}\frac{\partial}{\partial\xi^i})
\]

We can write an equation for a one-component field $\Phi(\xi)$ (an $HLM$ generalized scalar
field) with the help of the quadratic Casimir operator $C_2$, that is represented through the physical operators  $p_i$, $x_i$, $F_{ij}$, $i,j=0,...,3$, and $I$. Taking into account
the relations  (\ref{lin_tr}), we obtain \cite{khle}
\[
 (\sum_{i<j}F_{ij}F^{ij}(\frac{1}{M^{2}L^{2}}-
\frac{1}{H^{2}})+I^{2}+
\]
\begin{equation}
+\frac{x_{i}p^{i}+p_{i}x^{i}}{H}-
\frac{x_{i}x^{i}}{L^{2}}-\frac{p_{i}p^{i}}{M^{2}})\Phi(\xi)=0
\label{casim}
\end{equation}

\noindent   The equation (\ref{casim}) is similar to the Klein-Gordon-Fock equation of the canonical quantum theory. The explicit form of the  Casimir operator $C_2$, i.e. the left-hand side of the equation (\ref{casim}), was obtained in  Ref. \cite{khru20} at $H =\infty $ and  arbitrary values of $L$  and $M$. In a three-dimensional space the explicit form of
the quadratic Casimir operator $C_2$ was found in Ref. \cite{lezn}. 

\section{Groups of $LM$-generalized symmetries for quantum particles}
\label{ModelParameters}

One may choose kinematic groups with  specified properties taking into account
the properties of quantum particles and their interactions.
For instance, commutation relations (\ref{finalg}) at $H=\infty$ 
are suited for  symmetries of quantum particles, which conserve in interactions the
 $P$, $C$, and $T$ invariance. For example, this case 
 can be applied to strong interaction particles such as quarks \cite{khru20,b13}. Then  we  have the following commutation relations for $LM$ algebras in the system of 
 units with $\hbar=1$. 
\[
[F_{ij},F_{kl}] = 
i(g_{jk}F_{il}-g_{ik}F_{jl}+g_{il}F_{jk}-g_{jl}F_{ik}),
\]
\[
[F_{ij},p_k] = i(g_{jk}p_i - g_{ik}p_j),
\]
\[
[F_{ij},x_k] = i(g_{jk}x_i - g_{ik}x_j),
\]
\[
[p_i,x_j] = ig_{ij}I, 
\]
\begin{equation}
\label{stralg}
 [F_{ij},I] = 0,
\end{equation}
\[
[p_i,p_j] = (i/L^2)F_{ij},
\]
\[
 [x_i,x_j] = (i/M^2)F_{ij},
\]
\[
[p_i,I] = (i/L^2)x_i, 
\]
\[
[x_i,I] = (-i/M^2)p_i,
\]

\noindent 
In the limiting case  $L\to \infty $, $M\to \infty $  the canonical commutation relations (\ref{standart}) are gained. The equation of the second order for  $LM$ generalized fields have the following form:
\[
(\frac{1}{M^{2}L^{2}}\sum_{i<j}F_{ij}F^{ij}+I^{2}-
\]
\begin{equation}
-\frac{x_{i}x^{i}}{L^{2}}-\frac{p_{i}p^{i}}{M^{2}})\Phi(\xi)=0
\label{casimi}
\end{equation}

\section{Equations for LM-generalized spinor fields}
\label{LMfields}

Let us consider equations for  $LM$ generalized spinor fields which are invariant
in regards to transformations generated by algebras  (\ref{stralg}) \cite{kh-mea}.
The linear with respect to operators 
$p_i$, $x_j$, $F_{ij}$, and $I$ equations can be found with the help of the method
considered in Ref. \cite{iz94}. In the first place we write an equation for the generalized
field  $\psi(\xi)$ with four components: 
 \[
 (\gamma_ip^i - \gamma_i\gamma_5x^i\zeta_1\zeta_2\sqrt{-\frac{M^2}{L^2}}- \gamma_5I\zeta_2\sqrt{-M^2}-  
\]
\begin{equation}
 -\sum_{i<j}\gamma_i\gamma_jF^{ij}\frac{\zeta_1}{\sqrt{L^2}}-n 1_4)\psi(\xi)=0, 
 \label{nde}
\end{equation} 
\noindent where $F^{ij}$ are the Lorentz group generators,  
$n$ is a free parameter, $1_4$ is the unity $ 4\times 4$ matrix,
$\gamma_i$, $i=0,1,2,3$, are the Dirac matrices, $\zeta_1=\pm 1$, $\zeta_2=\pm 1$.

The equation (\ref{nde}) is not invariant with respect to spatial parity $P$. Below the 
$P$ invariant equation for the generalized
field  $\Psi(\xi)$ with eight  components is given, that is suitable for description of quantum particles with $P$ invariant interactions. 
\[
 (\sigma_0\otimes\gamma_ip^i 
-\sigma_3\otimes\gamma_5I\zeta_2\sqrt{-M^2}-
\]
\begin{equation}
- \sigma_3\otimes\gamma_i\gamma_5x^i\zeta_1\zeta_2\sqrt{-\frac{M^2}{L^2}}
 \label{ide}
\end{equation}
\[
-\sigma_0\otimes\sum_{i<j}\gamma_i\gamma_jF^{ij}\frac{\zeta_1}{\sqrt{L^2}}-\sigma_0\otimes n 1_4)\Psi(\xi)=0, 
\]  
\noindent where   $\sigma_0$ is the unity  $2\times 2$ matrix, $\sigma_3$ is
the Pauli matrix.
The coefficients in Eq.(5) obtained in Ref. \cite{b13} are specified with the explicit forms of the coefficients in Eqs.(\ref{nde}) and (\ref{ide}), which contain the discrete parameters 
 $\zeta_1$ and   $\zeta_2$. This permits to study in detail solutions of Eqs. (\ref{nde}) and (\ref{ide}).

\section{Concluding remarks}
\label{Section_Conclusion}

Equations (\ref{casim}), (\ref{nde}), and (\ref{ide}) for the  generalized scalar and spinor fields are  invariant with respect to the symmetries generated by the algebras with commutation relations (\ref{finalg}) or (\ref{stralg}), which are generalizations of canonical commutation relations (\ref{standart}). For this purpose some  finite-dimensional  and  infinite-dimensional representations of the generalized algebras are used. 
 
 Later on, it is possible to consider applications of the $HLM$  fields either in a domain of ultra-low distances in the quantum field theory with a generalized symmetry, as was assumed by Snyder, or for a model description of quantum gravity phenomena.  Moreover, 
at the present, the multi-messenger (photons, neutrinos, cosmic rays and gravitational waves) 
astronomy opens up the possibility of a search for phenomenological signatures of
quantum gravity in a space-time of such type \cite{adda}.
So it is useful to study in detail the peculiarities of solutions to the 
equations considered above  in generalized quantum variables spaces.

\end{document}